\documentclass[aps,prl,twocolumn,groupedaddress]{revtex4-1}
\usepackage{graphicx}

\begin{document}


\title{Constraining annihilating dark matter mass by the radio continuum spectral data of NGC4214 galaxy}


\author{Man Ho Chan, Chak Man Lee}
\affiliation{The Education University of Hong Kong, Tai Po, Hong Kong, China}


\date{\today}

\begin{abstract}
Recent gamma-ray and radio observations provide stringent constraints for annihilating dark matter. The current $2\sigma$ lower limits of dark matter mass can be constrained to $\sim 100$ GeV for thermal relic annihilation cross section. In this article, we use the radio continuum spectral data of a nearby galaxy NGC4214 and differentiate the thermal contribution, dark matter annihilation contribution and cosmic-ray contribution. We can get more stringent constraints of dark matter mass and annihilation cross sections. The $5\sigma$ lower limits of thermal relic annihilating dark matter mass obtained are 300 GeV, 220 GeV, 220 GeV, 500 GeV and 600 GeV for $e^+e^-$, $\mu^+\mu^-$, $\tau^+\tau^-$, $W^+W^-$ and $b\bar{b}$ channels respectively. These limits challenge the dark matter interpretation of the gamma-ray, positron and antiproton excess in our Milky Way.
\end{abstract}

\maketitle


\section{Introduction}
Recent observations of gamma-ray, positrons and antiprotons indicate some excess emissions of these particles in our Milky Way. These excess emissions could be explained by dark matter (DM) annihilation (i.e. the DM interpretations). For example, the gamma-ray excess can be explained by DM annihilating via $b\bar{b}$ channel with DM mass $m \sim 30-80$ GeV \cite{Daylan,Calore,Abazajian}. For positron excess and antiproton excess, the suggested mass is $m \sim 100-1000$ GeV \cite{Boudaud} and $m \approx 46-94$ GeV (via $b\bar{b}$ channel) \cite{Cholis} respectively. Except for the positron excess interpretation, the mass ranges coincide with each other and the annihilation cross sections predicted are close to the thermal relic annihilation cross section $<\sigma v>=2.2 \times 10^{-26}$ cm$^3$ s$^{-1}$ \cite{Steigman}. 

However, recent analyses of gamma-ray observations give very stringent constraints for annihilating DM. If DM particles are thermal relic particles (the simplest model in cosmology), the latest Fermi-LAT gamma-ray observations of Milky Way dwarf spheroidal satellite (MW dSphs) galaxies and two nearby galaxy clusters give the lower limits of DM mass $m \sim 100$ GeV for $b\bar{b}$ quark and $\tau^+\tau^-$ channels \cite{Ackermann,Albert,Chan}. For leptophilic channels like $e^+e^-$ and $\mu^+\mu^-$, analyses of AMS-02 data \cite{Bergstrom,Cavasonza} and radio data \cite{Egorov,Chan2,Chan3,Chan4} also give the lower limits $m \sim 50-90$ GeV. Some more recent analyses of radio data can improve the limits to $m \sim 300$ GeV \cite{Chan5}. These limits obtained give some tension to the DM interpretations of the gamma-ray, positron and antiproton excess. Nevertheless, most of these limits are only $2\sigma$ limits and it is still too early to rule out the possibility of the DM interpretations. 

In this article, we use the radio continuum spectral data of a nearby galaxy NGC4214 and differentiate the thermal contribution, dark matter annihilation contribution and cosmic-ray contribution. We show that the $5\sigma$ lower limits of DM mass with thermal relic annihilation cross section can be improved to $\ge 220$ GeV for leptophilic channels and $\ge 500$ GeV for two popular non-leptophilic channels. These results provide some challenges to the DM interpretations of the gamma-ray, positron and antiproton excess in our Milky Way.

\section{The model}
Previous studies show that radio data can give stringent constraints for annihilating DM \cite{Egorov,Chan2,Chan3,Chan4,Storm,Chan9,Chan5}. The high-energy electrons and positrons produced from DM annihilation emit strong synchrotron radiation $S_{DM}$ (in radio waves) when there is a strong magnetic field. These analyses assume that all radio fluxes emitted $S_{\rm total}$ originate from high-energy electrons and positrons produced from DM annihilation (i.e. $S_{\rm total}=S_{DM}$). This assumption overestimates the contribution of DM annihilation because high-energy electrons and positrons can also be produced by normal astrophysical processes such as supernovae and pulsars (normal cosmic rays). Therefore, the limits obtained for DM are somewhat underestimated.

If one can differentiate the contributions of radio flux emitted from a galaxy due to DM annihilation $S_{DM}$ and normal cosmic rays $S_{CR}$, the constraints of DM can be much more stringent. Furthermore, we can also eliminate the radio flux due to thermal contribution $S_{\rm th}$. The electromagnetic emission of the thermal electrons in a galaxy would contribute a small part in the radio flux. This thermal contribution part could be calculated by standard thermal physics \cite{Tabatabaei}. Including the thermal contribution can give more stringent limits for DM mass. In the following, we assume $S_{\rm total}=S_{DM}+S_{CR}+S_{\rm th}$. If the magnetic field of the galaxy is strong enough ($B \ge 5~\mu$G) and remains uniform in the outer region, the diffusion of high-energy electrons and positrons would be insignificant so that the radio flux (in mJy) contributed by DM emitted from a galaxy can be simply given by \cite{Profumo,Bertone}
\begin{equation}
S_{DM}(\nu) \approx \frac{1}{4 \pi \nu D^2} \left[ \frac{9 \sqrt{3} <\sigma v>}{2m^2(1+C)} E(\nu)Y(\nu,m) \int \rho_{DM}^2dV \right],
\end{equation}
where $\nu$ is the radio frequency, $<\sigma v>$ is the annihilation cross section, $D$ is the distance to the galaxy, $C$ is the correction factor for inverse Compton scattering contribution, $\rho_{DM}$ is the DM density profile of the galaxy, $E(\nu)=14.6(\nu/{\rm GHz})^{1/2}(B/{\rm \mu G})^{-1/2}$ GeV, $Y(\nu,m)=\int_{E(\nu)}^m(dN_e/dE')dE'$ and $dN_e/dE'$ is the energy spectrum of the electrons or positrons produced from DM annihilation (it depends on annihilation channels) \cite{Cirelli}. Here, we have used the `point-source approximation' in Eq.~(1). For angular region smaller than $1^{\circ}$, the `J-factor' is approximately equal to $J \approx \int \rho_{DM}^2dV/D^2$ \cite{Ullio}. As we will see below, the angular region of our target galaxy is much smaller than $1^{\circ}$ so that using the `point-source approximation' can be justified. Generally speaking, $S_{DM}(\nu)$ is a power-law of the radio frequency $\nu$ ($S_{DM}(\nu) \propto \nu^{-\alpha_{DM}}$). The spectral index of radio spectrum $\alpha_{DM}$ depends on different annihilation channels. For example, $\alpha_{DM} \approx 0.5$ for $e^+e^-$ channel while $\alpha_{DM}>1$ for $b\bar{b}$ channel.

On the other hand, numerical simulations show that the spectral index for GeV cosmic rays is very close to a constant (for $\nu \sim$ GHz) \cite{Nava}. This is also true for our Galaxy based on observations \cite{Drury}. In fact, many galaxies show nearly constant spectral index for a wide range of frequencies (e.g. NGC 4449, NGC 891) \cite{Srivastava,Mulcahy}, and even for galaxy clusters \cite{Murgia}. We will see later that the spectral index of our target galaxy is also constant. Therefore, assuming a constant spectral index for cosmic rays would give a better fit rather than the non-constant spectral index models. Otherwise, fine-tuning of the $S_{DM}$ and $S_{CR}$ is required to give a resultant constant spectral index for a wide range of frequencies. Also, the constant spectral index model is the simplest model for cosmic-ray emission (only two parameters are involved). Based on the above arguments and minimizing the involved parameters, we can write $S_{CR}=S_{CR,0}\nu^{-\alpha_{CR}}$. Our previous study using this model gives better constraints of DM in the Ophiuchus galaxy cluster \cite{Chan6}. For thermal contribution, we can write $S_{\rm th}=S_{\rm th,0}\nu^{-0.1}$ \cite{Srivastava}. The total radio flux emitted by a galaxy is given by
\begin{equation}
S_{\rm total}(\nu)=S_{DM}(\nu)+S_{CR,0}\nu^{-\alpha_{CR}}+S_{\rm th,0}\nu^{-0.1}.
\end{equation}
The parameter $S_{\rm th,0}$ can be obtained from observational data. Unfortunately, it is very difficult to predict the theoretical values of $S_{CR,0}$ and $\alpha_{CR}$ for a galaxy. These two values are free parameters when we apply this model to fit the observational data.

\section{The radio continuum spectral data}
There are some criteria to follow for choosing the best target galaxy for analysis. First of all, the galaxy chosen should have a large uniform magnetic field strength $B$. It is because a large uniform magnetic field strength can greatly facilitate the cooling of high-energy electrons and positrons produced from DM annihilation. Due to the high cooling rate, the diffusion of high-energy electrons and positrons would be insignificant and the resultant radio flux contributed by DM would be maximized. Second, the galaxy should be nearby and rich in DM content. Also, the radio data should have small uncertainties in both large and small frequency regimes. 

We have examined some archival galactic radio continuum spectral data and we have found a very good candidate - NGC 4214 galaxy - to do the analysis. The radio continuum data can be found in \cite{Srivastava}. The thermal contribution can be modeled by $S_{\rm th}=20(\nu/0.1~{\rm GHz})^{-0.1}$ mJy \cite{Srivastava}. Therefore, we can obtain the non-thermal radio flux data $S_{\rm nth}$ and their uncertainties (see Table 1). The non-thermal radio spectral index is very close to a constant so that it is very good for analysis. Note that the data in \cite{Srivastava} have included several observations from different telescopes. In particular, the data at $\nu=1.4$ GHz ($56.9 \pm 0.4$ mJy, $51.5 \pm 0.4$ mJy, $38.3 \pm 7.7$ mJy and $70 \pm 25$ mJy) and $\nu=4.85-4.86$ GHz  ($30.0 \pm 4.5$ mJy, $30.0 \pm 7.0$ mJy and $34.0 \pm 6.8$ mJy) have shown some discrepancies. Therefore, we combine the data at $\nu=1.4$ GHz and $\nu=4.85-4.86$ GHz respectively as $62.8 \pm 32.2$ mJy and $31.9 \pm 8.9$ mJy to allow for the largest possible observational uncertainties (see Table 1). 

The distance to the galaxy is $D=2.94$ Mpc \cite{Srivastava} and the average uniform magnetic field strength is $B \approx 8$ $\mu$G \cite{Kepley}. The angular size of the galaxy is smaller than $0.2^{\circ}$. Note that the radio flux data we considered are integrated flux which represent the total emissions of the galaxy. Since we do not have the information about the radio flux profile, we assume all radio signals come from a single halo with size smaller than $0.2^{\circ}$. As the size is smaller than $1^{\circ}$, the `point-source approximation' in Eq.~(1) is still a very good approximation. 

Generally speaking, the magnetic field of a galaxy usually trace the matter distribution and an exponential function is commonly assumed to model the magnetic field. However, this assumption requires two extra parameters (the central magnetic field and scale radius) and the functional form also contributes systematic uncertainties. In modeling radio emission of DM annihilation, a larger magnetic field would give a larger radio flux (except for the $e^+e^-$ channel) \cite{Chan2,Chan3}. Observations indicate that the central magnetic field of the NGC 4214 galaxy can be as high as $30~\mu$G and the magnetic field strength in the outer region (even for the outskirt region) is close to a uniform strength $8$ $\mu$G \cite{Kepley}. Therefore, the radio emission near the center is much larger. However, the actual central magnetic strength and the magnetic scale radius are quite uncertain. To avoid extra uncertain parameters involved, we adopt a `uniform field approximation' and use $B=8$ $\mu$G to model the magnetic field strength. This would underestimate the stronger radio emission due to DM annihilation. Using the constant magnetic field strength can give conservative limits of $m$, except for the $e^+e^-$ channel. Nevertheless, the effect of the assumption for the $e^+e^-$ channel is not very large (the limit of $m$ is larger by less than 40\%). 

Note that the magnetic field strength of NGC 4214 is somewhat higher than that in normal dwarf galaxies (e.g. Local Group dwarf galaxies: $B=4.2 \pm 1.8$ $\mu$G \cite{Chyzy}). Nevertheless, study in \cite{Chyzy} point out that some higher star-formation rate and starburst galaxies may have very high average magnetic field strength. For example, the magnetic field strengths of the NGC 2976 galaxy, NGC 1569 galaxy and NGC 4449 galaxy are $B=6.6 \pm 1.8$ $\mu$G, $B=14 \pm 3$ $\mu$G and $B=9 \pm 2$ $\mu$G respectively \cite{Drzazga,Chyzy}. Since the star-formation rate of NGC 4214 galaxy (SFR$=0.09M_{\odot}$/yr) is close to that of NGC 1569 (SFR$=0.13M_{\odot}$/yr) \cite{Hong}, the large magnetic field strength of the NGC 4214 galaxy ($B \approx 8$ $\mu$G) is not unexpected.

The effect of the inverse Compton scattering is also very important. Since the cooling rate of the inverse Compton scattering also depends on $E^2$, this effect can be simply characterized by a correction factor $C$ in Eq.~(1). Assuming a conservative optical-infrared radiation energy density $\omega_{\rm opt}=0.5$ eV/cm$^3$ for NGC 4214 (same as our Galaxy) \cite{Atoyan}, the energy density for inverse Compton scattering is about 0.75 eV/cm$^3$, which corresponds to $C \approx 0.49$. The synchrotron and the inverse Compton scattering are the dominated cooling processes.

As mentioned above, since the cooling rate is very high, the diffusion is not important in the NGC 4214 galaxy. This can be examined with the diffusion scale length \cite{Yuan}
\begin{equation}
\lambda \sim 0.79~{\rm kpc} \left( \frac{D_0}{\rm 10^{26}~cm^3} \right)^{1/2} \left( \frac{\omega_0}{\rm 1~eV~cm^{-3}} \right)^{-1/2} \left(\frac{E}{\rm 1~GeV} \right)^{-1/3},
\end{equation}
which represents the approximated length traveled by an electron with initial energy $E$. Here, $D_0$ is the diffusion coefficient and $\omega_0$ is the total radiation energy density. For $B=8$ $\mu$G and $C=0.49$, we have $\omega_0=2.3$ eV cm$^{-3}$. The value of the diffusion coefficient is scale-dependent. The standard diffusion coefficient used in our Milky Way galaxy is $D_0=3.1 \times 10^{28}$ cm$^2$ s$^{-1}$ \cite{Beck}. However, for a smaller NGC4214 galaxy, the smallest scale on which the magnetic field is homogeneous is somewhat smaller. The diffusion coefficient for a dwarf galaxy is of the order $10^{26}$ cm$^2$ s$^{-1}$ \cite{Jeltema}. For the injection spectrum, most of the electrons and positrons having $E \sim 1-100$ GeV for $m \ge 100$ GeV, which correspond to $\lambda \sim 0.1-1$ kpc. This means that an electron with $E=1-100$ GeV would lose most of its energy by traveling a distance of $0.1-1$ kpc. Since most of the high-energy electrons and positrons are produced near the central region of the galaxy (because of the much higher density), most of them would lose all their energy and they would be confined within the galaxy (the size of the galaxy is larger than 5.63 kpc). Note that the central magnetic field of the NGC 4214 is much stronger. The diffusion scale length is much shorter for the dominant central emission. 

The DM density profile of NGC 4214 can be probed from the SPARC data \cite{Lelli}. The SPARC data include the observed rotational velocity $v$ and the rotational velocity contributed by baryonic matter $v_b$. By subtracting the baryonic matter contribution and assuming a standard value of mass-to-luminosity ratio for galaxies $\Upsilon=0.5M_{\odot}/L_{\odot}$ \cite{Lelli}, we can obtain the rotational velocity contributed by DM $v_{DM}^2=v^2-v_b^2$. The DM density can be calculated by $\rho_{DM}=(4\pi r^2)^{-1}(d/dr)(rv_{DM}^2/G)$. In Fig.~1, we can see that the resulting DM density can be well-fitted by a power-law form $\rho_{DM} \propto r^{-2.18}$ for $r>1.67$ kpc. For the central density within $r \le 1.67$ kpc, the uncertainties are quite large and we assume a constant density profile which can give a conservative prediction of DM annihilation signal. Therefore, we model the DM density as
\begin{equation}
\rho_{DM}= \left \{ \begin{array}{ll}
\rho_0       & {\ \ r \le 1.67~{\rm kpc} } \\ &\\
\rho_0 \left(\frac{r}{\rm 1.67~kpc} \right)^{-2.18 \pm 0.06}       & {\ \  1.67~{\rm kpc}<r \le 5.63~{\rm kpc} } \end{array} \right. ,
\end{equation}
where $\rho_0=(3.4 \pm 0.8) \times 10^{-24}$ g cm$^{-3}$. The uncertainty of fitting is very small. Here, we do not assume any particular forms of DM density profile (e.g. Navarro-Frenk-White profile or Burkert profile). The DM density profile is just directly probed from the observed rotation curve data. The systematic uncertainties involved would be smaller than assuming any particular forms of DM density profile.

Simulations show that the DM annihilation signal would be enhanced due to the substructure contributions. These contributions can be quantified by considering the boost factor $B_f$, which can be modeled by the following empirical expression \cite{Moline}:
\begin{equation}
\log B_f= \sum_{i=0}^5b_i \left( \log \frac{M}{M_{\odot}} \right)^i,
\end{equation}
where $M$ is the virial mass of the structure and $b_i$ is the fitted coefficients \cite{Moline}. Following the DM profile in Eq.~(4), the virial mass is $M=9.8 \times 10^{10}M_{\odot}$. Using the most conservative model in \cite{Moline}, the corresponding boost factor is $B_f=4.44$. 

We follow the standard cosmological scenario and assume the thermal relic annihilation cross section for DM $<\sigma v>=2.2 \times 10^{-26}$ cm$^3$ s$^{-1}$ \cite{Steigman}. Therefore, we can get $S_{DM}(\nu)$ for different annihilation channels and different DM mass $m$. For each annihilation channel and $m$, we can fit the predicted $S_{DM}(\nu)+S_{CR}(\nu)$ with the non-thermal radio continuum spectral data of NGC4214 $S_{\rm nth}$ obtained in \cite{Srivastava}. We minimize the reduced $\chi^2$ value ($\chi_{\rm red}^2$) by changing the values of two free parameters, $S_{CR,0}$ and $\alpha_{CR}$. 

In Table 2, we present the corresponding $\chi_{\rm red}^2$ values for some DM mass and annihilation channels. The $5\sigma$ lower limits of $m$ are 300 GeV, 220 GeV, 220 GeV, 500 GeV and 600 GeV for $e^+e^-$, $\mu^+\mu^-$, $\tau^+\tau^-$, $W^+W^-$ and $b\bar{b}$ channels respectively, which are determined by the relation between $\chi_{\rm red}^2$ and $m$ in Fig.~2. We also plot the spectra for $m$ just ruled out at $5\sigma$ and just satisfied the $2\sigma$ lower limits respectively in Fig.~3. The corresponding components of the thermal contribution, DM contribution and the cosmic-ray contribution are shown in Fig.~4. In particular, we notice that for the $e^+e^-$, $\mu^+\mu^-$ and $\tau^+\tau^-$ channels, the best-fit scenarios do not have the contributions of cosmic rays ($S_{CR}=0$, see Fig.~4 and Table 2). It means that the spectral index of dark matter annihilation for these three channels are already very close to the observed non-thermal radio spectrum so that no cosmic-ray component is needed to give better fits. In other words, dark matter contribution alone plus thermal component is sufficient to give the best-fit spectra for these three channels.

In fact, this is the first time that we can rule out $m \le 220$ GeV at $5\sigma$ for thermal relic annihilating DM. Generally speaking, the $\chi_{\rm red}^2$ will decrease further and finally approach to a constant if we increase the value of $m$ (see Fig.~2). It is because the non-thermal radio continuum spectrum of NGC 4214 is very close to a constant spectral index $\alpha_{nth}=-0.63 \pm 0.04$ \cite{Srivastava}. Increasing the value of $m$ would suppress the contribution of $S_{DM}$ so that $S_{CR} \approx S_{\rm nth}$. Therefore, we can only obtain the lower limits of $m$ using this method. 

In the above analysis, we take the value of the thermal relic annihilation cross section $<\sigma v>=2.2 \times 10^{-26}$ cm$^3$ s$^{-1}$. Nevertheless, DM particles may not be thermal relic particles and the annihilation cross section may be larger or smaller than the thermal relic annihilation cross section. If we release the annihilation cross section as a free parameter, we can obtain its upper limit for each annihilation channel. However, the values of the annihilation cross section and the two free parameters, $S_{CR,0}$ and $\alpha_{CR}$, are quite degenerate for a particular value of $m$. Therefore, we fix the values of $S_{CR}$ and $\alpha_{CR}$ as their convergent limits (the best-fit values when $m$ is very large) and obtain the $5\sigma$ upper limits of the annihilation cross section as a function of $m$ (see Fig.~5). We also show the $2\sigma$ upper limits of the annihilation cross section obtained by the Fermi-LAT gamma-ray observations of the MW dSphs galaxies \cite{Ackermann,Albert} in Fig.~5. We can see that our $5\sigma$ upper limits are tighter than the $2\sigma$ Fermi-LAT gamma-ray limits.

We also examine the lower limits of DM if there is no cosmic-ray contribution. Generally speaking, if the cosmic-ray contribution is zero ($S_{CR}=0$), the lower limits of $m$ would be smaller because the DM contribution has to be larger to account for the radio spectrum (smaller $m$ gives larger $S_{DM}$). Therefore, setting $S_{CR}=0$ would give the most conservative lower limits of $m$. However, the thermal component $S_{\rm th}$ is determined by the H$\alpha$ emission measurement \cite{Srivastava}, which is independent of the radio observations. Therefore, the thermal component cannot be set to zero arbitrarily. In Fig.~6, we show the $\chi_{\rm red}^2$ values for the 5 channels without cosmic-ray contributions. We can see that the $5\sigma$ DM mass ranges for the $e^+e^-$, $\mu^+\mu^-$ and $\tau^+\tau^-$ channels are 300-540 GeV (best-fit: 360 GeV), 220-400 GeV (best-fit: 280 GeV) and 220-430 GeV (best-fit: 280 GeV) respectively. For the $b\bar{b}$ and $W^+W^-$ channels, the $5\sigma$ ranges of $m$ are 600-1050 GeV (best-fit: 800 GeV) and 490-830 GeV (best-fit: 600 GeV) respectively. The resulting $5\sigma$ lower limits of $m$ are nearly the same as the results including cosmic-ray contributions (compare with the results in Table 2). That means the dark matter contribution is still dominating at these lower limits of $m$. However, the best-fit $\chi_{\rm red}^2$ for the $b\bar{b}$ and $W^+W^-$ channels are larger than 4, which are excluded at more than $3.8\sigma$. Therefore, excluding the cosmic-ray contribution gives poorer fits (larger $\chi_{\rm red}^2$ values) for these two channels and better fits will be obtained if cosmic-ray contributions are included (compare the $\chi_{\rm red}^2$ in Table 2). 

Note that the $\chi_{\rm red}^2$ values in Fig.~2 approach to a small constant value while the $\chi_{\rm red}^2$ increase with $m$ in Fig.~6 in the large DM mass regime. It is because the calculations of $\chi_{\rm red}^2$ in Fig.~2 have included the cosmic-ray component. The cosmic-ray component would dominate the contribution in the large DM mass regime and make the $\chi_{\rm red}^2$ values small. The best-fit $\chi_{\rm red}^2$ value is about 1.5 without DM contribution (only cosmic-ray and thermal contributions), which is a very good fit indeed. In other words, the cosmic-ray and thermal contributions alone can give a very good explanation for the radio continuum data of the NGC 4214. A large contribution of the DM component (when $m$ is sufficiently small) would give a large value of $\chi_{\rm red}^2$. That's why we can obtain the lower limits of $m$ by using the radio continuum data of NGC 4214.

\begin{figure}
\vskip 10mm
 \includegraphics[width=85mm]{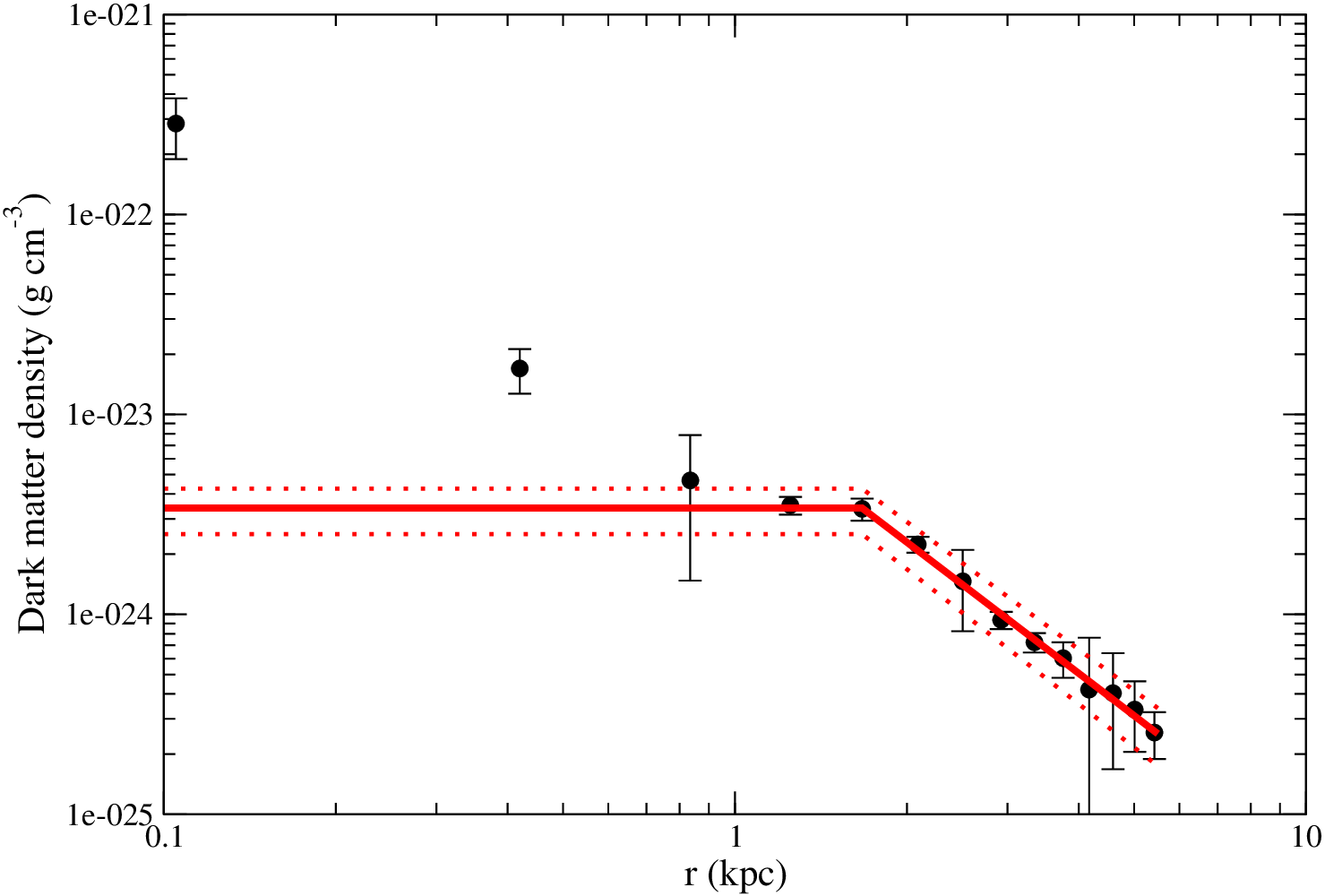}
 \caption{The DM density profile probed from the rotation curve data in \cite{Lelli}. The red solid line is the density model in Eq.~(4) and the red dotted lines indicate the $1 \sigma$ uncertainty.}
\vskip 10mm
\end{figure}

\begin{figure}
\vskip 10mm
 \includegraphics[width=85mm]{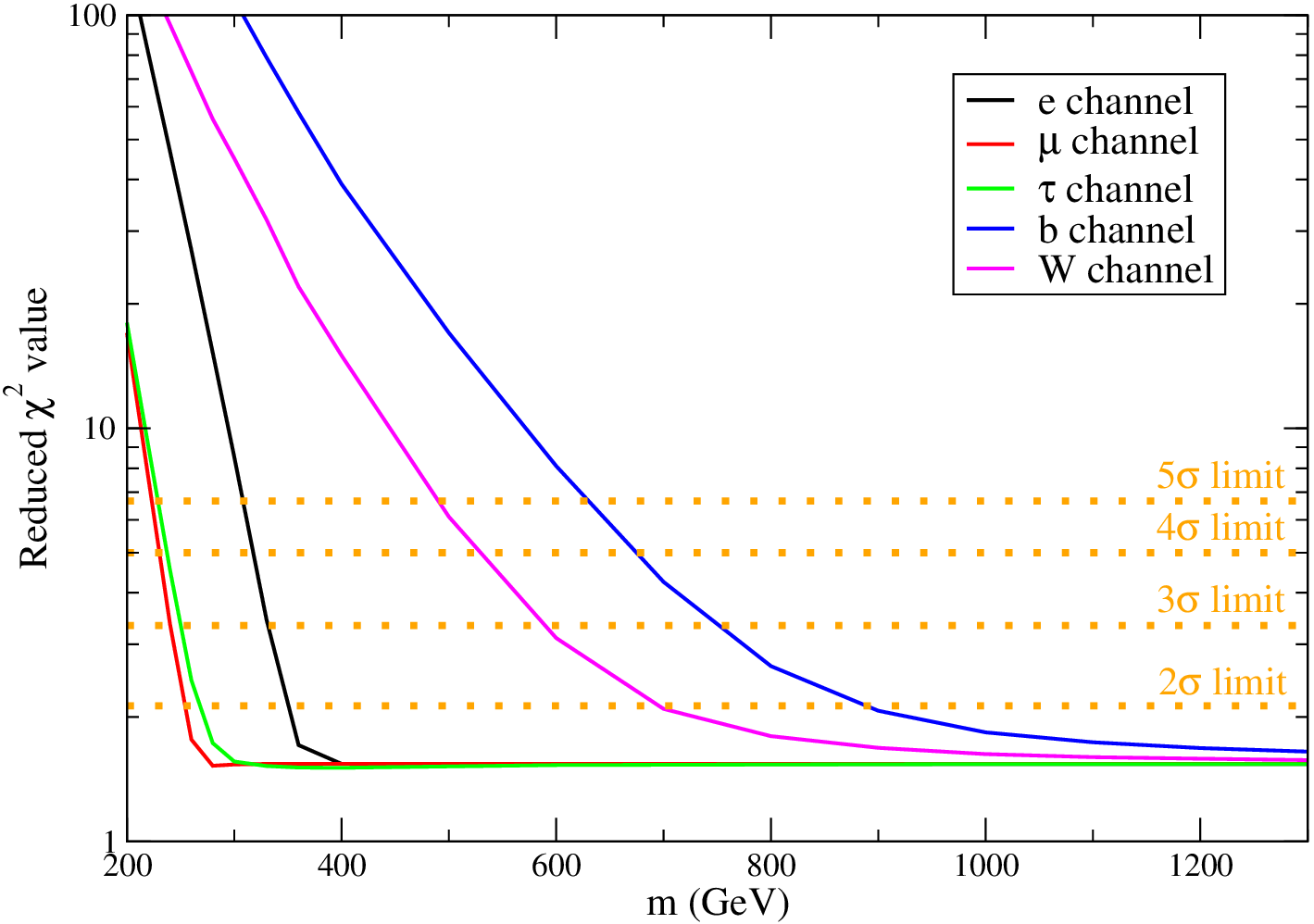}
 \caption{The relation between the reduced $\chi^2$ values and the DM mass $m$ for various channels.}
\vskip 10mm
\end{figure}

\begin{figure}
\vskip 10mm
 \includegraphics[width=85mm]{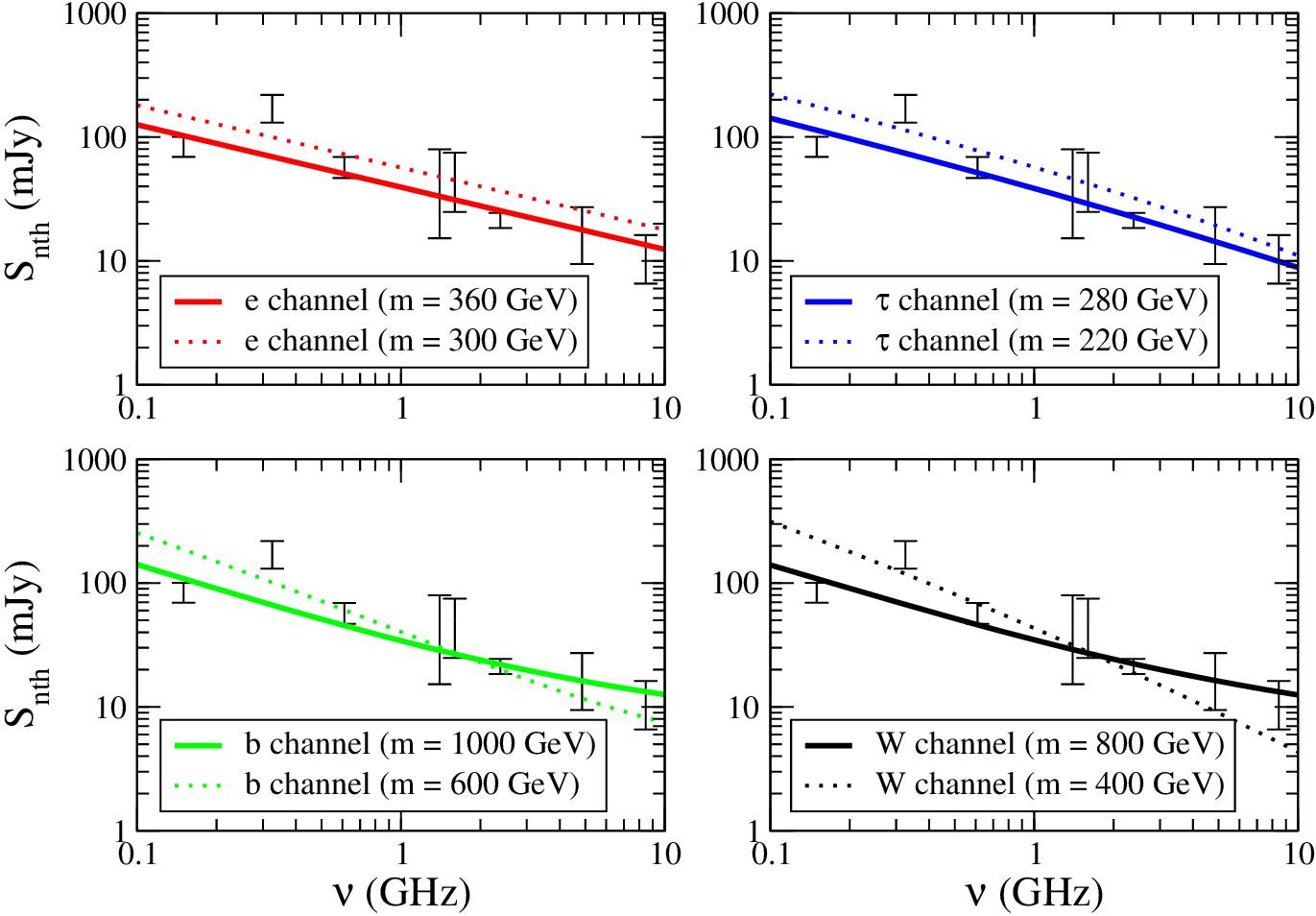}
 \caption{The best-fit spectra of $S_{\rm nth}$ for $m$ just ruled out at $5\sigma$ (dotted lines) and just satisfied the $2\sigma$ lower limits (solid lines). The corresponding best-fit parameters and the reduced $\chi^2$ values are shown in Table 2.}
\vskip 10mm
\end{figure}

\begin{figure}
\vskip 10mm
 \includegraphics[width=85mm]{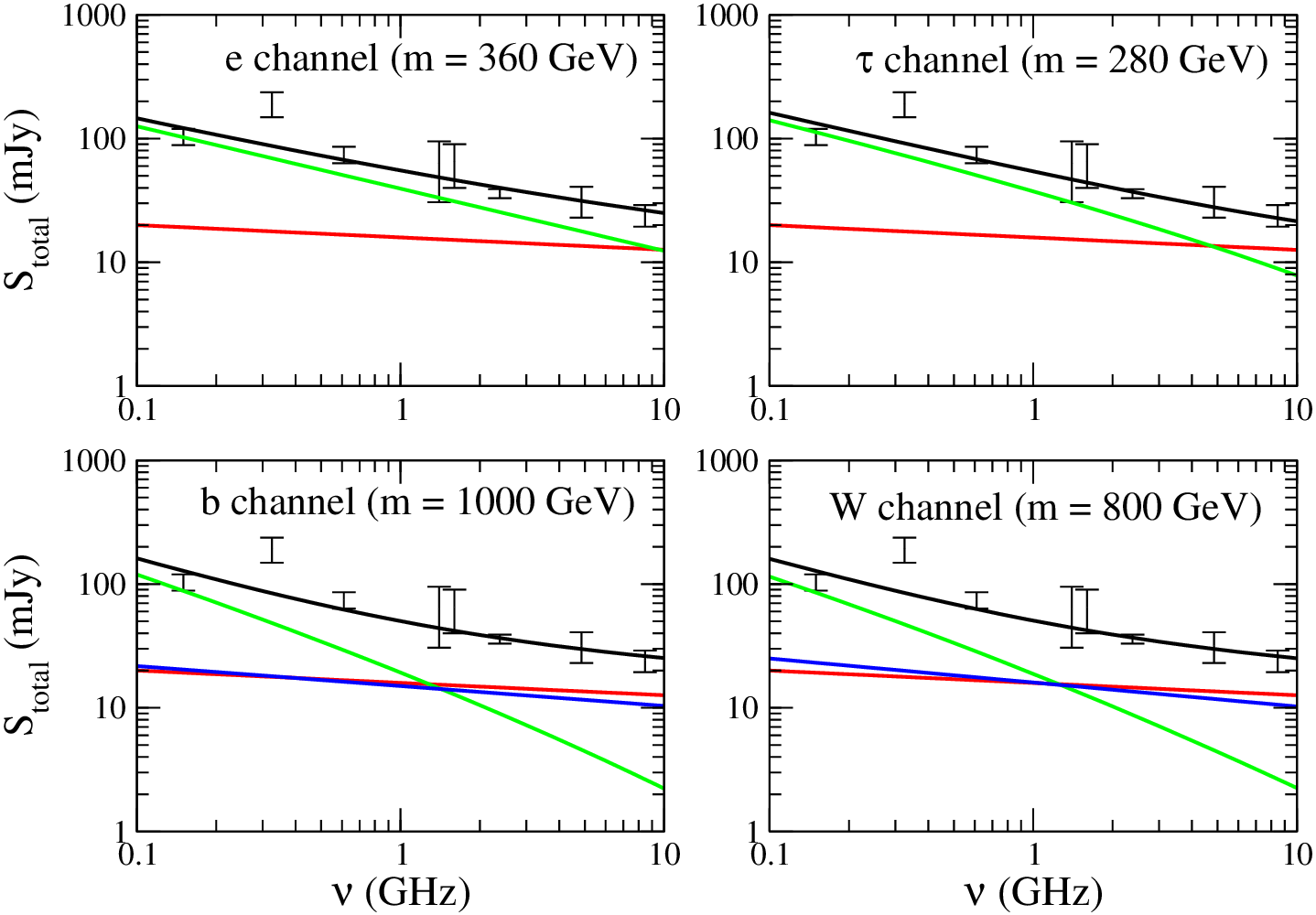}
 \caption{The best-fit spectra of $S_{\rm total}$ and the corresponding components (for $m$ just satisfied the $2\sigma$ lower limits). The black lines, red lines, green lines and blue lines indicate the total radio flux $S_{\rm total}$, thermal radio flux contribution $S_{\rm th}$, DM flux contribution $S_{DM}$ and cosmic-ray flux contribution $S_{CR}$ respectively. The corresponding best-fit parameters are shown in Table 2.}
\vskip 10mm
\end{figure}

\begin{figure}
\vskip 10mm
 \includegraphics[width=85mm]{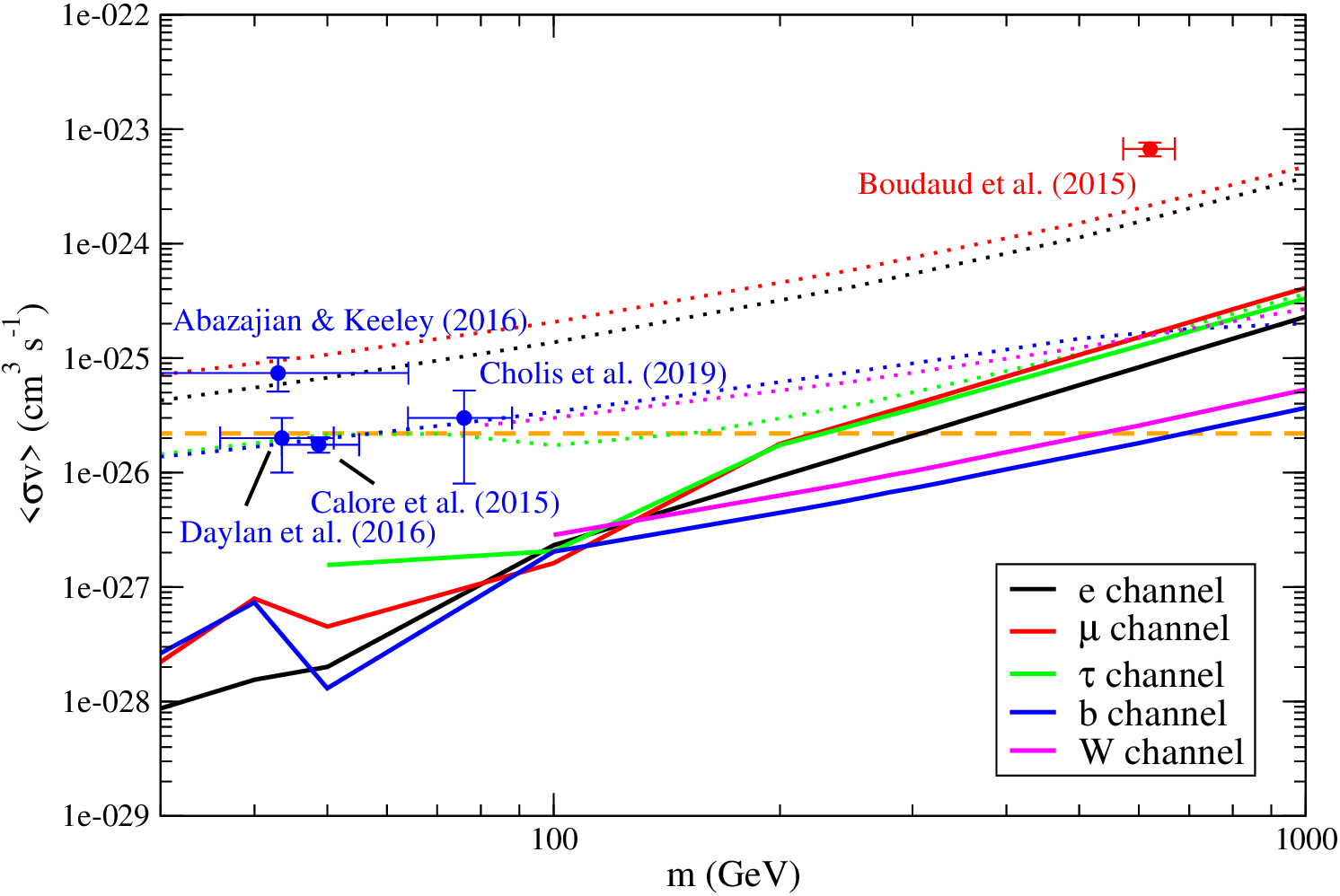}
 \caption{The colored solid lines are the $5\sigma$ upper limits of the annihilation cross section in our analysis. The colored dotted lines are the $2\sigma$ upper limits of the annihilation cross section obtained by the Fermi-LAT gamma-ray observations of the MW dSphs galaxies \cite{Ackermann,Albert}. The orange dashed line indicates the thermal relic annihilation cross section $<\sigma v>=2.2\times 10^{-26}$ cm$^3$ s$^{-1}$ \cite{Steigman}. The data with error bars shown are the predicted ranges of $m$ and $<\sigma v>$ based on  the dark matter interpretations of gamma-ray excess \cite{Calore,Daylan,Abazajian}, positron excess \cite{Boudaud} and antiproton excess \cite{Cholis}. The blue and red colors of the data points correspond to the ranges for the $b\bar{b}$ and $\mu^+\mu^-$ channels respectively.}
\vskip 10mm
\end{figure}

\begin{figure}
\vskip 10mm
 \includegraphics[width=85mm]{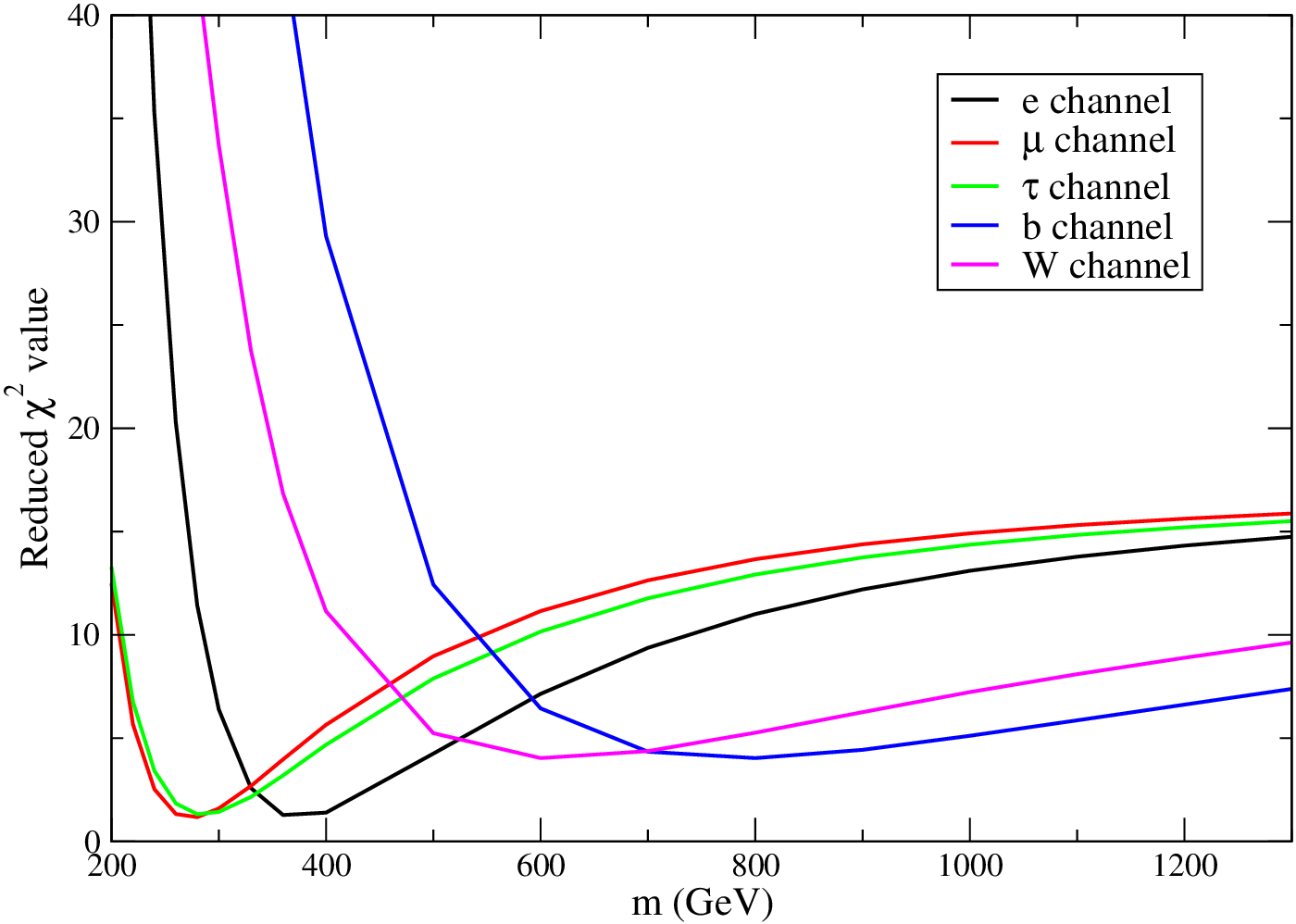}
 \caption{The relation between the reduced $\chi^2$ values and the DM mass $m$ for various channels without cosmic-ray contributions.}
\vskip 10mm
\end{figure}

\begin{table}
\caption{The radio continuum spectral data of NGC 4214 \cite{Srivastava}. The four data for $\nu=1.4$ GHz and three data for $\nu=4.855$ GHz shown in \cite{Srivastava} have been combined correspondingly to allow for the largest possible uncertainties.}
 \label{table1}
 \begin{tabular}{@{}lccc}
  \hline
  $\nu$ (GHz) &  $S_{\rm total}$ (mJy) & $S_{\rm nth}$ (mJy) & Uncertainties (mJy) \\
  \hline
  0.15 & 104.1 & 84.9 & 15.6 \\
  0.325 & 192.5 & 174.7 & 43.9 \\
  0.61 & 74.6 & 57.9 & 11.2 \\
  1.4 & 62.8 & 47.4 & 32.2 \\
  1.6 & 65 & 49.8 & 25 \\
  2.38 & 36 & 21.4 & 3 \\
  4.855 & 31.9 & 18.3 & 8.9 \\
  8.46 & 24.2 & 11.4 & 4.8 \\
  \hline
 \end{tabular}
\end{table}

\begin{table}
\caption{The best fit parameters for some DM mass $m$ and annihilation channels.}
 \label{table2}
 \begin{tabular}{@{}lccccc}
  \hline
  Channel & $m$ (GeV) &  $\chi_{\rm red}^2$ & $S_{CR,0}$ (mJy) & $\alpha_{CR}$ & Remark \\
  \hline
   & 300 & 8.53 & 0 & 0 & Ruled out at $5\sigma$ \\
  $e^+e^-$ & 330 & 3.45 & 0 & 0 & Ruled out at $3\sigma$ \\
  & 360 & 1.71 & 0 & 0 & Within $2\sigma$ range \\
  \hline
  & 220 & 7.56 & 0 & 0 & Ruled out at $5\sigma$ \\
  $\mu^+\mu^-$ & 240 & 3.37 & 0 & 0 & Ruled out at $3\sigma$ \\
  & 260 & 1.77 & 0 & 0 & Within $2\sigma$ range \\
  \hline
  & 220 & 9.02 & 0 & 0 & Ruled out at $5\sigma$ \\
  $\tau^+\tau^-$ & 260 & 2.45 & 0 & 0 & Ruled out at $2\sigma$ \\
  & 280 & 1.73 & 0 & 0 & Within $2\sigma$ range \\
  \hline  
   & 400 & 14.9 & 0 & 0 & Ruled out at $5\sigma$ \\
  $W^+W^-$ & 600 & 3.11 & 9 & 0 & Ruled out at $2\sigma$ \\
   & 800 & 1.80 & 16 & 0.19 & Within $2\sigma$ range \\
  \hline
   & 600 & 8.10 & 4 & 0 & Ruled out at $5\sigma$ \\
  $b\bar{b}$ & 800 & 2.66 & 9 & 0 & Ruled out at $3\sigma$ \\
   & 1000 & 1.84 & 15 & 0.16 & Within $2\sigma$ range \\
  \hline
 \end{tabular}
\end{table}

\section{Discussion}
In this article, we use the radio continuum spectrum of a galaxy (NGC 4214) to obtain the lower limits of DM mass $m$ for five popular annihilation channels. Using radio data is a very good option for constraining DM because current radio telescopes can give observations with very high resolution and sensitivity. For the NGC 4214 radio data we used, the radio beam size and flux density level detected can be as small as 5" and 1 mJy respectively \cite{Srivastava}. Therefore, the radio data obtained may be more effective in constraining annihilating DM than the Milky Way gamma-ray or positron data used in previous studies. Furthermore, we have differentiated the contributions of the thermal emissions, DM annihilation and the normal cosmic rays so that we can obtain a better lower limit of $m$ for each of the annihilation channels. In fact, many recent studies of gamma rays and positrons have included the astrophysical background components \cite{Belikov,Gammaldi,Cembranos}. Nevertheless, using appropriate radio continuum spectral data with the consideration of the background cosmic-ray and thermal components seem to get better constraints. The limits obtained are the current most stringent radio limits for thermal relic annihilating DM, which challenge the DM interpretations of the gamma-ray excess \cite{Daylan,Calore,Abazajian} and antiproton excess \cite{Cholis}. For the positron excess, it requires a much larger annihilation cross section ($\ge 10^{-24}$ cm$^3$ s$^{-1}$) \cite{Boudaud}. Our results also rule out the proposed DM interpretation if we assume $<\sigma v> \ge 10^{-24}$ cm$^3$ s$^{-1}$ (see Fig.~5). If we do not consider the boost factor, the lower limits of $m$ would approximately decrease by a factor of 2.3. Therefore, the minimum $5\sigma$ limits of $m$ is still larger than 90 GeV for all popular channels. This still challenges the DM interpretations of the positron and gamma-ray excess. In fact, the DM interpretations of the gamma-ray and positron excess are controversial. The ranges of DM mass and annihilation cross sections predicted are close to our expected values while some other studies point out that the excess emissions might originate from pulsars or molecular clouds \cite{Linden,Bartels,Boer}. Our results may provide some hints for settling this controversy.

Note that the above results are solely based on the data of a single galaxy. In fact, the diffusion processes and cosmic-ray emissions in a small galaxy are not very well known. For instance, if the diffusion length of the high-energy electrons and positrons is much longer than our expected, the radio emission due to the DM contribution would be suppressed and the resulting lower limits of DM mass would be smaller. Therefore, our results may be affected by the systematic uncertainties involved. More observations and analysis using a larger sample of galaxies are definitely required to examine and verify our claims.

The advantage of using the radio continuum spectral data is that the spectral index is close to a constant. This is true for many galaxies and galaxy clusters \cite{Drury,Srivastava,Murgia,Mulcahy}. Therefore, this method can be applied in many good targets (nearby DM-rich galaxies) to constrain DM. More radio continuum observations for these galaxies are definitely helpful. This method can also be applied in analyzing galaxy clusters \cite{Chan6,Chan8}. Using appropriate target objects, the $5\sigma$ limits of DM mass could be improved to $500-1000$ GeV. We expect that this method can open the `TeV window' for DM, which is complementary to other analyses of high-energy observations such as H.E.S.S. \cite{Abdallah} and DAMPE mission \cite{Ambrosi}. Future radio detection by the Square Kilometre Array (SKA) may be able to constrain thermal relic DM mass up to 10 TeV \cite{Cembranos2}. In fact, the null result of DM direct-detection experiments \cite{Tan,Aprile} may suggest that DM mass is of the order or larger than $\sim 1$ TeV \cite{Cooley}. Some recent analyses of the DAMPE data indicate that DM mass may be even larger than 1.4 TeV \cite{Jin,Chan7}. Nonetheless, many theoretical proposals suggest $m \sim 100-1000$ GeV \cite{Roszkowski,Bertone2}. Therefore, if we can improve the $5\sigma$ constraints to $\sim 500-1000$ GeV before we could have a more sensitive detector, it can help rule out most of the existing DM models and narrow down the possible parameter space for DM. 

\section{acknowledgements}
The work described in this paper was supported by a grant from the Research Grants Council of the Hong Kong Special Administrative Region, China (Project No. EdUHK 28300518) and the internal research fund from the Education University of Hong Kong (RG2/2019-2020R).

\end{document}